\documentclass[epj]{svjour}
\usepackage{graphics}
\begin{document}
\title{Chiral perturbation theory}
\subtitle{Success and challenge}
\author{Stefan Scherer
}                     
%
%
\institute{Institut f\"ur Kernphysik, Johannes Gutenberg-Universit\"at Mainz,
J.~J.~Becher Weg 45, D-55099 Mainz, Germany}
\date{Received: date / Revised version: date}
%
\abstract{Chiral perturbation theory is the effective field theory of the strong
interactions at low energies.
   We will give a short introduction to chiral perturbation theory for mesons and
will discuss, as an example, the electromagnetic polarizabilities of the pion.
   These have recently been extracted from an experiment on radiative $\pi^+$
photoproduction from the proton ($\gamma p\to \gamma \pi^+ n$) at the Mainz
Microtron MAMI.
   Next we will turn to the one-baryon sector of chiral perturbation
theory and will address the issue of a consistent power counting scheme.
   As examples of the heavy-baryon framework we will comment on the extraction
of the axial radius from pion electroproduction and will discuss the generalized
polarizabilities of the proton.
   Finally, we will discuss two recently proposed manifestly Lorentz-invariant
renormalization schemes and illustrate their application in a calculation of the
nucleon electromagnetic form factors.
} 
\maketitle
\section{Introduction}
\label{intro} Chiral perturbation theory (ChPT)
\cite{Weinberg:1978kz,Gasser:1983yg,Gasser:1984gg,Gasser:1987rb} is the effective
field theory (EFT) \cite{Ecker:2005ny} of the strong interactions at low
energies. The central idea of the EFT approach was formulated by Weinberg as
follows \cite{Weinberg:1978kz}: ``...  if one writes down the most general
possible Lagrangian, including {\em all} terms consistent with assumed symmetry
principles, and then calculates matrix elements with this Lagrangian to any given
order of perturbation theory, the result will simply be the most general possible
S--matrix consistent with analyticity, perturbative unitarity, cluster
decomposition and the assumed symmetry principles.''
   In the context of the strong interactions these ideas have first been applied
to the interactions among the Goldstone bosons of spontaneous symmetry breaking
in quantum chromodynamics (QCD).
   The effective theory is formulated in terms of the asymptotically observed states
instead of the quark and gluon degrees of freedom of the underlying (fundamental)
theory, namely QCD.
   The corresponding EFT---mesonic chiral perturbation theory---has been
tested at the two-loop level (see, e.g., \cite{Scherer:2002tk,Scherer:2005ri} for
a pedagogical introduction).
   A successful EFT program requires both the knowledge of the most general
Lagrangian up to and including the given order one is interested in as well as an
expansion scheme for observables.
   Due to the vanishing of the Goldstone boson masses in the chiral limit
in combination with their vanishing interactions in the zero-energy limit, a
derivative and quark-mass expansion is a natural scenario for the corresponding
EFT.
   At present, in the mesonic sector the Lagrangian is known up to and including
${\cal O}(q^6)$, where $q$ denotes a small quantity such as a four momentum or a
pion mass.
   The combination of dimensional regularization with the modified
minimal subtraction scheme of ChPT \cite{Gasser:1983yg} leads to a
straightforward correspondence between the loop expansion and the chiral
expansion in terms of momenta and quark masses at a fixed ratio, and provides a
consistent power counting for renormalized quantities.

   In the extension to the one-nucleon sector \cite{Gasser:1987rb} an additional
scale, namely the nucleon mass, enters the description.
   In contrast to the Goldstone boson masses, the nucleon mass does not vanish
in the chiral limit.
   As a result, the straightforward correspondence between the loop expansion and the
chiral expansion of the mesonic sector, at first sight, seems to be lost:
higher-loop diagrams can contribute to terms as low as ${\cal O}(q^2)$
\cite{Gasser:1987rb}.
   This problem has been eluded in the framework of the
heavy-baryon formulation of ChPT \cite{Jenkins:1990jv,Bernard:1992qa}, resulting
in a power counting analogous to the mesonic sector.
   The basic idea consists in expressing the relativistic nucleon field
in terms of a velocity-dependent field, thus dividing nucleon momenta into a
large piece close to on-shell kinematics and a soft residual contribution.
   Most of the calculations in the one-baryon sector have been performed
in this framework (for an overview see, e.g., \cite{Bernard:1995dp}) which
essentially corresponds to a simultaneous expansion of matrix elements in $1/m$
and $1/(4\pi F_\pi)$.
   However, there is price one pays when giving up manifest Lorentz invariance of the
Lagrangian.
   At higher orders in the chiral expansion, the
expressions due to the $1/m$ corrections of the Lagrangian become increasingly
complicated \cite{Ecker:1995rk,Fettes:2000gb}.
   Moreover, not all of the scattering amplitudes, evaluated perturbatively
in the heavy-baryon framework, show the correct analytical behavior in the
low-energy region \cite{Bernard:1996cc}.
   In recent years, there has been a
considerable effort in devising  renormalization schemes leading to a simple and
consistent power counting for the renormalized diagrams of a manifestly
Lorentz-invariant approach
\cite{Tang:1996ca,Ellis:1997kc,Becher:1999he,Lutz:1999yr,%
Gegelia:1999gf,Gegelia:1999qt,Lutz:2001yb,Fuchs:2003qc}.

   In the following we will highlight a few topics in chiral perturbation
theory which have been subject of experimental tests at the Mainz Microtron MAMI.

\section{Chiral perturbation theory for mesons}
\label{chptmesons}
\subsection{The effective Lagrangian and Weinberg's power counting scheme}
\label{elwpc}
The starting point of mesonic chiral perturbation theory is a
chiral $\mbox{SU}(N_l)_L\times\mbox{SU}(N_l)_R\times \mbox{U}(1)_V$ symmetry of
the QCD Lagrangian for $N_l$ massless (light) quarks:
\begin{equation}
\label{lqcd0lr} {\cal L}^0_{\rm QCD}=\sum_{l=1}^{N_l}
(\bar{q}_{R,l}iD\hspace{-.6em}/\hspace{.3em}q_{R,l}+\bar{q}_{L,l}iD
\hspace{-.6em}/\hspace{.3em} q_{L,l})-\frac{1}{4}{\cal G}_{\mu\nu,a} {\cal
G}^{\mu\nu}_a.
\end{equation}
   In eq.~(\ref{lqcd0lr}), $q_{L,l}$ and $q_{R,l}$ denote the left- and
right-handed components of the light quark fields.
   Here, we will be concerned with the cases $N_l=2$ and $N_l=3$ referring to
massless $u$ and $d$ or $u$, $d$ and $s$ quarks, respectively.
   Furthermore, we will
neglect the terms involving the heavy quark fields.
   The covariant derivative $D_\mu q_{L/R,l}$ contains the flavor-independent
coupling to the eight gluon gauge potentials, and ${\cal G}_{\mu\nu,a}$ are the
corresponding field strengths.
   The Lagrangian of eq.\ (\ref{lqcd0lr}) is invariant
under separate global $\mbox{SU}(N_l)_{L/R}$ transformations of the left- and
right-handed fields.
   In addition, it has an overall $\mbox{U}(1)_V$ symmetry.
   Several empirical facts give rise to the assumption that this chiral symmetry
is spontaneously broken down to its vectorial subgroup
$\mbox{SU}(N_l)_V\times\mbox{U}(1)_V$.
   For example, the low-energy hadron spectrum seems to follow multiplicities
of the irreducible representations of the group $\mbox{SU}(N_l)$ (isospin SU(2)
or flavor SU(3), respectively) rather than
$\mbox{SU}(N_l)_L\times\mbox{SU}(N_l)_R$, as indicated by the absence of
degenerate multiplets of opposite parity.
   Moreover, the lightest mesons form a pseudoscalar octet with masses that
are considerably smaller than those of the corresponding vector mesons.
   According to Coleman's theorem \cite{Coleman:1966}, the symmetry pattern of the
spectrum reflects the invariance of the vacuum state.
   Therefore, as a result of Goldstone's theorem \cite{Goldstone:1961eq,Goldstone:1962es},
one would expect $6-3=3$ or $16-8=8$ massless Goldstone bosons for $N_l=2$ and
$N_l=3$, respectively.
   These Goldstone bosons have vanishing interactions as their energies tend
to zero.
   Of course, in the real world, the pseudoscalar meson multiplet is not
massless which is a result of the finite quark masses of the $u$, $d$ and $s$
quarks.
   This explicit symmetry breaking in terms of the quark masses is treated
   perturbatively.

   The symmetries as well as the symmetry breaking pattern of QCD---once the quark masses
   are included---are mapped
onto the most general (effective) Lagrangian for the interaction of the Goldstone
bosons.
      The Lagrangian is organized in the number of the (covariant) derivatives
and of the quark mass terms
\cite{Weinberg:1978kz,Gasser:1983yg,Gasser:1984gg,Issler:1990nj,%
Akhoury:1990px,Scherer:1994wi,Fearing:1994ga,Bijnens:1999sh,Ebertshauser:2001nj,Bijnens:2001bb}
\begin{equation}
\label{lpi} {\cal L}_\pi={\cal L}_2 + {\cal L}_4 + {\cal L}_6 + \cdots,
\end{equation}
where the lowest-order Lagrangian is given by\footnote{In the following, we will
give equations for the two-flavor case.}
\begin{equation}
\label{l2} {\cal L}_2 = \frac{F^2}{4} \mbox{Tr} \Big [ D_{\mu} U
(D^{\mu}U)^{\dagger} +\chi U^{\dagger}+ U \chi^{\dagger} \Big ].
\end{equation}
Here,
\begin{displaymath}
U(x)=\exp\left(i\frac{\phi}{F}\right),\quad \phi=\left(\begin{array}{cc} \pi^0
&\sqrt{2}\pi^+\\ \sqrt{2}\pi^-&-\pi^0\end{array}\right),
\end{displaymath}
is a unimodular unitary $(2\times 2)$ matrix containing the Goldstone boson
fields.
   In eq.~(\ref{l2}), $F$ denotes the pion-decay constant in the chiral limit:
$F_\pi=F[1+{\cal O}(\hat{m})]=92.4$ MeV.
   When including the electromagnetic interaction, the covariant derivative
is defined as $D_\mu U = \partial_\mu U +ie A_\mu [Q,U],$ where
$Q=\mbox{diag}(2/3,-1/3)$ denotes the quark charge matrix.
   We work in the isospin-symmetric limit $m_u=m_d=\hat{m}$.
The quark masses are contained in $\chi=2 B \hat{m} =M^2$, where $M^2$ denotes
the lowest-order expression for the squared pion mass and $B$ is related to the
quark condensate $\langle \bar{q} q\rangle_0$ in the chiral limit.
   The next-to-leading-order Lagrangian contains 7 low-energy constants $l_i$
\cite{Gasser:1983yg}
\begin{eqnarray}
\label{lag4} \lefteqn{{\cal L}_4=l_5\left[\mbox{Tr}(f^R_{\mu\nu} U f_L^{\mu\nu}
U^\dagger) -\frac{1}{2}\mbox{Tr}(f_{\mu\nu}^L f_L^{\mu\nu} +f_{\mu\nu}^R
f_R^{\mu\nu}) \right]}\nonumber\\
&&\hspace{-.5em}+i\frac{l_6}{2}\mbox{Tr}\left(f^R_{\mu\nu}D^\mu U(D^\nu
U)^\dagger +f^L_{\mu\nu}(D^\mu U)^\dagger D^\nu U\right)+\cdots,
\end{eqnarray}
where we have displayed those terms which will be relevant for the discussion of
Compton scattering below.
   In that case, the field strength is given by
$$f^R_{\mu\nu}=f^L_{\mu\nu}=-e
(\partial_\mu A_\nu-\partial_\nu A_\mu)Q.$$

   In addition to the most general Lagrangian, one needs a method to assess
the importance of various diagrams calculated from the effective Lagrangian.
  Using Weinberg's power counting scheme \cite{Weinberg:1978kz}
one may analyze the behavior of a given diagram calculated in the framework of
eq.~(\ref{lpi}) under a linear re-scaling of all {\em external} momenta,
$p_i\mapsto t p_i$, and a quadratic re-scaling of the light quark masses, $\hat
m\mapsto t^2 \hat m$, which, in terms of the Goldstone boson masses, corresponds
to $M^2\mapsto t^2 M^2$.
   The chiral dimension $D$ of a given diagram with amplitude
${\cal M}(p_i,\hat m)$ is defined by
\begin{equation}
\label{mr1} {\cal M}(tp_i, t^2 \hat m)=t^D {\cal M}(p_i,\hat m),
\end{equation}
where, in $n$ dimensions,
\begin{eqnarray}
D&=&n N_L-2I_\pi+\sum_{k=1}^\infty 2k N_{2k}^\pi\label{mr2a}\\
&=&2+(n-2) N_L+\sum_{k=1}^\infty 2(k-1) N_{2k}^\pi \label{mr2b}\\
&\geq&\mbox{2 in 4 dimensions}.\nonumber
\end{eqnarray}
   Here, $N_L$ is the number of independent loop momenta,
$I_\pi$ the number of internal pion lines, and $N_{2k}^\pi$ the number of
vertices originating from ${\cal L}_{2k}$.
   A diagram with chiral dimension $D$ is said to be of order ${\cal O}(q^D)$.
   Clearly, for small enough momenta and masses diagrams with small $D$, such
as $D=2$ or $D=4$, should dominate.
   Of course, the re-scaling of eq.~(\ref{mr1}) must be viewed as
a mathematical tool.
   While external three-momenta can, to a certain extent, be made arbitrarily
small, the re-scaling of the quark masses is a theoretical instrument only.
   Note that, for $n=4$, loop diagrams are always suppressed due to the term
$2N_L$ in eq.~(\ref{mr2b}).
   In other words, we have a perturbative scheme in terms of external
momenta and masses which are small compared to some scale (here $4\pi F\approx 1$
GeV).

\begin{figure}
\begin{center}
\resizebox{0.2\textwidth}{!}{%
\includegraphics{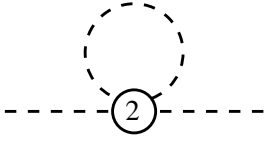}
}
\end{center}
\caption{One-loop contribution to the pion self-energy. The number 2 in the
interaction blob refers to ${\cal L}_2$.} \label{fig:examplepionselfenergy}
\end{figure}

\begin{figure}
\begin{center}
\resizebox{0.2\textwidth}{!}{%
\includegraphics{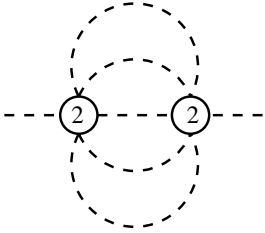}
}
\end{center}
\caption{Four-loop contribution to the pion self-energy.} \label{fig:exampleres3}
\end{figure}
    Figures \ref{fig:examplepionselfenergy} and \ref{fig:exampleres3} show contributions to the
pion self-energy with $D=4\cdot 1-2\cdot 1+ 2 \cdot 1=4$ and $D=4\cdot 4 -2\cdot
5+2\cdot 2 = 10$, respectively.
   As a specific example, let us consider the
contribution of fig.~\ref{fig:examplepionselfenergy} to the pion self-energy.
   Without going
into the details, the explicit result of the one-loop contribution is given by
(see, e.g., \cite{Scherer:2002tk})
\begin{displaymath}
\Sigma_{\rm loop}(p^2)= \frac{4p^2-M^2}{6 F^2} I_\pi(M^2,\mu^2,n) ={\cal O}(q^4),
\end{displaymath}
   where the dimensionally regularized integral is given by
 \begin{equation}
\label{I} I_\pi(M^2,\mu^2,n)=\frac{M^2}{16\pi^2}\left[
R+\ln\left(\frac{M^2}{\mu^2}\right)\right]+O(n-4).
\end{equation}
In eq.~(\ref{I}), $R$ is defined as
\begin{equation}
\label{R} R=\frac{2}{n-4}-[\mbox{ln}(4\pi)-\gamma_E+1],
\end{equation}
   with $n$ denoting the number of space-time dimensions and
$\gamma_E=-\Gamma'(1)$ being Euler's constant.
   Note that both factors---the
fraction and the integral---each count as ${\cal O}(q^2)$ resulting in ${\cal
O}(q^4)$ for the total expression as anticipated.
   In other words, when calculating one-loop graphs, using vertices from ${\cal L}_2$ of
eq.~(\ref{l2}), one generates infinities (so-called ultraviolet divergences).
      In the framework of dimensional regularization
these divergences appear as poles at space-time dimension $n=4$, since $R$ is
infinite as $n\to 4$.
   The loop diagrams are renormalized by absorbing the infinite parts
into the redefinition of the fields and the parameters of the most general
Lagrangian.
   Since ${\cal L}_2$ of eq.~(\ref{l2}) is not renormalizable in
the traditional sense, the infinities cannot be absorbed by a renormalization of
the coefficients $F$ and $B$.
   However, to quote from ref.\ \cite{Weinberg:1995mt}:
   ``... the cancellation of ultraviolet divergences does
not really depend on renormalizability; as long as we include every one of the
infinite number of interactions allowed by symmetries, the so-called
non-renormalizable theories are actually just as renormalizable as renormalizable
theories.''
   According to Weinberg's power counting of eq.~(\ref{mr2b}),
one-loop graphs with vertices from ${\cal L}_2$ are of ${\cal O}(q^4)$.
   The conclusion is that one needs to adjust (renormalize) the parameters
of ${\cal L}_4$ to cancel one-loop infinities.
   In doing so, one still has the freedom of choosing a suitable
renormalization condition.
   For example, in the minimal subtraction scheme (MS) one would fix the
parameters of the counterterm Lagrangian such that they would precisely absorb
the contributions proportional to $2/(n-4)$ .
   In the modified minimal subtraction scheme of ChPT
($\widetilde{\rm MS}$) employed in \cite{Gasser:1983yg}, the seven (bare)
coefficients $l_i$ of the ${\cal O}(q^4)$ Lagrangian of (\ref{lag4}) are
expressed in terms of renormalized coefficients $l_i^r$ as
\begin{equation}
\label{li} l_i=l_i^r+\gamma_i\frac{R}{32\pi^2},
\end{equation}
where the $\gamma_i$ are fixed numbers.

\subsection{Electromagnetic polarizabilities of the pion}
\label{sec:2}
   In the framework of classical electrodynamics, the electric and magnetic
polarizabilities $\alpha$ and $\beta$ describe the response of a system to a
static, uniform, external electric and magnetic field in terms of induced
electric and magnetic dipole moments.
   In principle, empirical information on the pion polarizabilities can be obtained
from the differential cross section of low-energy Compton scattering on a charged
pion
\begin{eqnarray*}
\lefteqn{\frac{d\sigma}{d\Omega_{lab}}= \left(\frac{\omega'}{\omega}\right)^2
\frac{e^2}{4\pi M_\pi}\left\{\frac{e^2}{4\pi M_\pi}
\frac{1+z^2}{2}\right.}\\
&&\left.-\frac{\omega\omega'}{2} \left[(\alpha+\beta)_{\pi^+}(1+z)^2
+(\alpha-\beta)_{\pi^+}(1-z)^2\right]\right\}\\
&& +\cdots,
\end{eqnarray*}
where $z=\hat{q}\cdot\hat{q}\,'$ and $\omega'/\omega=[1+\omega(1-z)/M_\pi]$.
   The forward and backward differential cross sections are
sensitive to $(\alpha+\beta)_{\pi^+}$ and $(\alpha-\beta)_{\pi^+}$, respectively.

   The predictions for the charged pion polarizabilities at ${\cal O}(q^4)$
\cite{Bijnens:1987dc} result from an old current-algebra low-energy theorem
\cite{Terentev:1972ix}
\begin{eqnarray*}
\alpha_{\pi^+}=-\beta_{\pi^+}&=&2 \frac{e^2}{4\pi} \frac{1}{(4\pi F_\pi)^2
M_\pi}\frac{\bar l_6-\bar l_5}{6}\\[1em]
&=&(2.64\pm 0.09) \times 10^{-4}\, \mbox{fm}^3,
\end{eqnarray*}
which relates Compton scattering on a charged pion, $\gamma\pi^+\to\gamma\pi^+$,
in terms of a chiral Ward identity to radiative charged-pion beta decay,
$\pi^+\to e^+\nu_e\gamma$.
  The linear combination $\bar l_6-\bar l_5$ of scale-independent low-energy constants
\cite{Gasser:1983yg} is fixed using the most recent determination of the ratio of
the pion axial-vector form factor $F_A$ and the vector form factor $F_V$ via the
radiative pion beta decay \cite{Frlez:2003pe}:
\begin{displaymath}
\gamma=\frac{1}{6}(\bar l_6-\bar l_5)=\frac{F_A}{F_V}=0.443\pm 0.015.
\end{displaymath}
   A two-loop analysis (${\cal O}(q^6)$) of the charged-pion
polarizabilities has been worked out in
\cite{Burgi:1996mm,Burgi:1996qi}\footnote[2]{Ref.\
\cite{Burgi:1996mm,Burgi:1996qi} uses $(\bar l_6-\bar l_5)=2.7\pm 0.4$ instead of
$2.64\pm 0.72$ which was obtained in ref.\ \cite{Gasser:1983yg} from
$\gamma=0.44\pm 0.12$. Correspondingly, this also generates a smaller error in
the ${\cal O}(q^4)$ prediction $\alpha_{\pi^+}=(2.7\pm 0.4) \times 10^{-4}\,
\mbox{fm}^3$ instead of $(2.62\pm 0.71)\times 10^{-4}\, \mbox{fm}^3$.}:
\begin{eqnarray}
\label{alphaplusbetachptop6}
(\alpha+\beta)_{\pi^+}&=&(0.3 \pm 0.1) \times 10^{-4}\, \mbox{fm}^3,\\
\label{alphaminusbetachptop6} (\alpha-\beta)_{\pi^+}&=&(4.4\pm 1.0) \times
10^{-4}\, \mbox{fm}^3.
\end{eqnarray}
    The degeneracy $\alpha_{\pi^+}=-\beta_{\pi^+}$ is lifted at the two-loop
level.
   The corresponding corrections amount to an 11\% (22\%) change of
the ${\cal O}(q^4)$ result for $\alpha_{\pi^+}$ ($\beta_{\pi^+}$), indicating a
similar rate of convergence as for the $\pi\pi$-scattering lengths
\cite{Gasser:1983yg,Bijnens:1995yn}.
   The effect of the new low-energy constants appearing at ${\cal O}(q^6)$
on the pion polarizability was estimated via resonance saturation by including
vector and axial-vector mesons.
   The contribution was found to be about 50\% of the two-loop result.
   However, one has to keep in mind that \cite{Burgi:1996mm,Burgi:1996qi} could
not yet make use of the improved analysis of radiative pion decay which, in the
meantime, has also been evaluated at two-loop accuracy
\cite{Bijnens:1996wm,Geng:2003mt}.
   Taking higher orders in the quark mass expansion into account,
Bijnens and Talavera obtain $(\bar l_6-\bar l_5)=2.98\pm 0.33$
\cite{Bijnens:1996wm}, which would slightly modify the leading-order prediction
to $\alpha_{\pi^+}=(2.96\pm 0.33)\times 10^{-4}\, \mbox{fm}^3$ instead of
$\alpha_{\pi^+}=(2.7\pm 0.4) \times 10^{-4}\, \mbox{fm}^3$ used in
\cite{Burgi:1996mm,Burgi:1996qi}.
   Accordingly, the difference $(\alpha-\beta)_{\pi^+}$ of
(\ref{alphaminusbetachptop6}) would increase to $4.9\times 10^{-4}\,\mbox{fm}^3$
instead of $4.4\times 10^{-4}\, \mbox{fm}^3$, whereas the sum would remain the
same as in eq.~(\ref{alphaplusbetachptop6}).

   As there is no stable pion target, empirical information about the pion
polarizabilities is not easy to obtain.
   For that purpose, one has to consider reactions which contain
the Compton scattering amplitude as a building block, such as, e.g., the
Primakoff effect in high-energy pion-nucleus bremsstrahlung, $\pi^-Z\to \pi^-Z
\gamma$ \cite{Antipov:1982kz}, radiative pion photoproduction on the nucleon,
$\gamma p\to \gamma \pi^+n$ \cite{Aibergenov:1986gi,Ahrens:2004mg}, and pion pair
production in $e^+e^-$ scattering, $e^+e^-\to e^+e^-\pi^+\pi^-$
\cite{Berger:1984xb,Courau:1986gn,Ajaltoni,Boyer:1990vu}.
   The results of the older experiments are summarized in table
   \ref{tab:pionpolexp}.

\begin{table}
\caption{Previous experimental data on the charged pion polarizability $\alpha_{\pi^+}$.}
\label{tab:pionpolexp}
\begin{tabular}{llc}
\hline\noalign{\smallskip} Reaction&Experiment&$\alpha_{\pi^+}$ [$10^{-4}$ $\mbox{fm}^3$]\\
\noalign{\smallskip}\hline\noalign{\smallskip} $\pi^{-}Z\rightarrow
\pi^{-}Z\gamma$ & Serpukhov \cite{Antipov:1982kz}& $6.8\pm 1.4\pm 1.2$\\
$\gamma p\to \gamma\pi^+ n$& Lebedev Phys.~Inst.~\cite{Aibergenov:1986gi}&$20\pm 12$\\
$\gamma \gamma \rightarrow \pi^+\pi^-$ & PLUTO \cite{Berger:1984xb}& $19.1\pm 4.8\pm 5.7$\\
& DM 1 \cite{Courau:1986gn}& $17.2\pm 4.6$      \\
& DM 2 \cite{Ajaltoni}& $26.3\pm 7.4$      \\
& MARK II \cite{Boyer:1990vu}& $2.2\pm 1.6$     \\
\noalign{\smallskip}\hline
\end{tabular}
\end{table}

\begin{figure}
\begin{center}
\resizebox{0.2\textwidth}{!}{%
\includegraphics{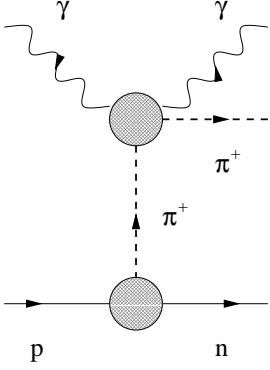}
}
\end{center}
\caption{The reaction $\gamma p\to\gamma\pi^+n$ contains Compton scattering on
a pion as a sub diagram in the $t$ channel, where $t=(p_n-p_p)^2$.}
\label{fig:tchannel}
\end{figure}

\begin{figure}
\begin{center}
\resizebox{0.4\textwidth}{!}{%
\includegraphics{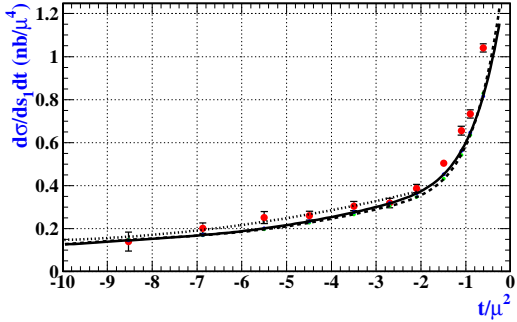}
}
\end{center}
\caption{Differential cross section averaged over 537 MeV $< E_\gamma <$ 817 MeV
and 1.5 $M_\pi^2<s_1<5 M_\pi^2$. Solid line: model 1; dashed line: model 2;
dotted line: fit to experimental data.} \label{fig:cros1c}
\end{figure}

   The potential of studying the influence of the pion polarizabilities on radiative pion
photoproduction from the proton was extensively studied in
\cite{Drechsel:1994kh}.
   In terms of Feynman diagrams, the reaction $\gamma p\to\gamma\pi^+n$ contains
real Compton scattering on a charged pion as a pion pole diagram (see
fig.~\ref{fig:tchannel}).
   In the recent experiment on $\gamma p\to\gamma\pi^+n$
at the Mainz Microtron MAMI \cite{Ahrens:2004mg}, the cross section was obtained
in the kinematic region 537 MeV $< E_\gamma <$ 817 MeV,
$140^{\circ}\le\theta^{\rm cm}_{\gamma\gamma'}\leq 180^{\circ}$.
   The values of the pion polarizabilities have been obtained from a
fit of the cross section calculated by different theoretical models to the data
rather than performing an extrapolation to the $t$-channel pole of the Chew-Low
type \cite{Chew:1958wd,Unkmeir}.
   Figure \ref{fig:cros1c} shows the
experimental data, averaged over the full photon beam energy interval and over
the squared pion-photon center-of-mass energy $s_1$ from 1.5 $M_\pi^2$ to 5
$M_\pi^2$ as a function of the squared pion momentum transfer $t$ in units of
$M_\pi^2$.
   For such small values of $s_1$, the differential cross section is expected to
be insensitive to the pion polarizabilities.
   Also shown are two model calculations: model 1 (solid curve) is a simple Born
approximation using the pseudoscalar pion-nucleon interaction including the
anomalous magnetic moments of the nucleon; model 2 (dashed curve) consists of
pole terms without the anomalous magnetic moments but including contributions
from the resonances $\Delta (1232)$, $P_{11}(1440)$, $D_{13}(1520)$ and
$S_{11}(1535)$.
   The dotted curve is a fit to the experimental data.

\begin{figure}
\begin{center}
\resizebox{0.4\textwidth}{!}{%
\includegraphics{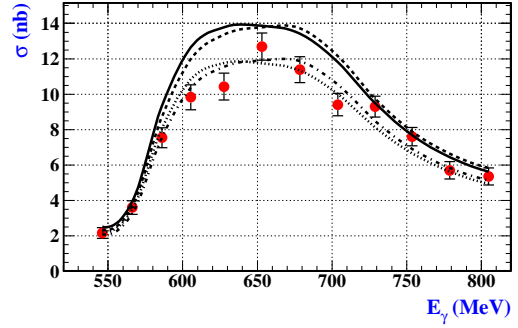}
}
\end{center}
\caption{
The cross section of the process $\gamma p\to\gamma\pi^+n$ integrated over $s_1$
and $t$ in the region where the contribution of the pion polarizability is
biggest and the difference between the predictions of the theoretical models
under consideration does not exceed 3 \%. The dashed and dashed-dotted lines are
predictions of model 1 and the solid and dotted lines of model 2 for
$(\alpha-\beta)_{\pi^+}=0$ and $(\alpha-\beta)_{\pi^+}=14\times 10^{-4}\,
\mbox{fm}^3$, respectively.} \label{fig:cros2c}
\end{figure}

   The kinematic region where the polarizability contribution
is biggest is given by $5M_\pi^2< s_1<15M_\pi^2$ and $-12M_\pi^2<t<-2M_\pi^2$.
   Figure \ref{fig:cros2c} shows the cross section as a function of the beam
energy integrated over $s_1$ and $t$ in this second region.
   The dashed and solid lines (dashed-dotted and dotted lines) refer to
models 1 and 2, respectively, each with $(\alpha-\beta)_{\pi^+}=0$
($(\alpha-\beta)_{\pi^+}=14\times 10^{-4}\, \mbox{fm}^3$).
   By comparing the experimental data of the 12 points with the predictions of
the models, the corresponding values of $(\alpha-\beta)_{\pi^+}$ for each data
point have been determined in combination with the corresponding statistical and
systematic errors.
    The result extracted from the combined analysis of the 12 data points
reads \cite{Ahrens:2004mg}
\begin{equation}
\label{alphambeta} (\alpha-\beta)_{\pi^+}=(11.6\pm 1.5_{\rm stat}\pm 3.0_{\rm
syst}\pm 0.5_{\rm mod}) \times 10^{-4}\, \mbox{fm}^3
\end{equation}
and has to be compared with the ChPT result of, say,  $(4.9\pm 1.0) \times
10^{-4}\, \mbox{fm}^3$ which deviates by 2 standard deviations from the
experimental result.
   On the other hand, the application of dispersion sum rules as performed in
\cite{Fil'kov:1998np} yields $(\alpha-\beta)_{\pi^+}=(10.3\pm 1.9) \times
10^{-4}\, \mbox{fm}^3$.

   Both the precision measurement of radiative pion beta decay \cite{Frlez:2003pe}
and of radiative pion photoproduction indicate that further theoretical and
experimental work is needed. In particular, the analysis of ref.\
\cite{Frlez:2003pe} suggests an inadequacy of the present $V-A$ description of
the radiative beta decay, which would also reflect itself in an inadequacy of the
ChPT description in its present form.
   What remains to be understood is why the dispersion sum rules give such
a dramatically different result from the ChPT calculation where the higher-order
terms have been estimated from resonance saturation by including vector and
axial-vector mesons.
   Clearly, the model-dependent input deserves further study. In this context, a full and
consistent one-loop calculation of $\gamma p\to\gamma\pi^+n$ including the Delta
resonance \cite{Hacker:2005fh} would be desirable.

   For a discussion of the so-called generalized pion polarizabilities see
\cite{Unkmeir:1999md,Fuchs:2000pn,L'vov:2001fz,Unkmeir:2001gw}.

\subsection{Future perspectives at MAMI}
   With the setup of the Crystal Ball detector, a dedicated $\eta$ physics program
will be possible at MAMI.
   In the reaction $\gamma + p \to p +\eta$, $10^7$ etas will be produced per
day.
   The main physics objectives will be the investigation of neutral decay
channels.

   In the framework of $\mbox{SU(3)}_L\times\mbox{SU(3)}_R$ symmetry the decay process
$\eta\to\pi^0\gamma\gamma$ is closely related to $\gamma\gamma\to\pi^0\pi^0$.
   At ${\cal O}(q^4)$, the amplitude is given entirely in terms of one-loop
   diagrams involving vertices of ${\cal O}(q^2)$.
   The prediction for the decay width was found to be two orders of magnitude
smaller \cite{Ametller:1991dp} than the measured value.
   The pion loops are small due to approximate $G$-parity invariance whereas the
   kaon loops are suppressed by the large kaon mass in the propagator.
   Therefore, higher-order contributions must play a dominant role in
$\eta\to\pi^0\gamma\gamma$.
   Even at ${\cal O}(q^6)$ differences of a factor of two are found for the
decay rate and spectrum
\cite{Ametller:1991dp,Ko:1993rg,Bellucci:1995ay,Bel'kov:1995fj,%
Jetter:1995js,Bijnens:1995vg,Oset:2002sh} although the most recent result for the
decay width of $\Gamma(\eta\to\pi^0\gamma\gamma)=(0.45\pm 0.12)$ eV agrees with
the original prediction $(0.42 \pm 0.20)$ eV of ref.~\cite{Ametller:1991dp}.
   The
decay $\eta\to\pi^0\pi^0\pi^0$ is a sensitive test of isospin symmetry violation
with the transition amplitude being proportional to the light quark mass
difference $(m_u-m_d)$ \cite{Gasser:1984pr,Leutwyler:1996tz}.
   Moreover, the electromagnetic interaction was shown to produce only a small
   contribution \cite{Baur:1995gc}.
As a final example for ``allowed'' decays we refer to the rare eta decay
$\eta\to\pi ^0\pi^0\gamma\gamma$ \cite{Knochlein:1996ah,Nefkens:2005ka}.
   On the other hand, in the forbidden decays such as $\eta\to\pi^0\pi^0$
and $\eta\to 4\pi^0$ one will investigate (P,CP) violation which may be connected
to the so-called $\theta$ term in QCD.

   As a final example we would like to point at the potential of investigating
the $\gamma \pi^+\to\pi^+\pi^0$ amplitude in the $\gamma p\to n \pi^+\pi^0$
reaction.
   This would allow for an alternative test of the Wess Zumino Witten action
\cite{Wess:1971yu,Witten:1983tw} in terms of the ${\cal F}_{3\pi}$ amplitude (see
\cite{Giller:2005uy} for a recent overview).

\section{Chiral perturbation theory for baryons}

\subsection{The power counting problem}
   The standard effective Lagrangian relevant to the single-nucleon sector
contains, in addition to eq.\ (\ref{lpi}), the most general $\pi N$ Lagrangian
\cite{Gasser:1987rb,Ecker:1995rk,Fettes:2000gb},
\begin{equation}
\label{lpin} {\cal L}_{\pi N}={\cal L}_{\pi N}^{(1)}+{\cal L}_{\pi
N}^{(2)}+\cdots.
\end{equation}
   Due to the additional spin degree of freedom ${\cal L}_{\pi N}$
contains both odd and even powers in small quantities.
   In order to illustrate the issue of power counting in the baryonic sector,
we consider the lowest-order $\pi N$ Lagrangian \cite{Gasser:1987rb}, expressed
in terms of bare fields and parameters denoted by subscripts 0,
\begin{equation}
\label{lpin1} {\cal L}_{\pi N}^{(1)}=\bar \Psi_0 \left( i\gamma_\mu
\partial^\mu -m_0 -\frac{1}{2}\frac{{\stackrel{\circ}{g_{A}}}_0}{F_0}
\gamma_\mu \gamma_5 \tau^a \partial^\mu \pi^a_0\right) \Psi_0 +\cdots,
\end{equation}
   where $\Psi_0$ and $\vec{\pi}_0$ denote a doublet and a triplet of bare
nucleon and pion fields, respectively.
   After renormalization, $m$, $\stackrel{\circ}{g_A}$, and $F$ refer to the
chiral limit of the physical nucleon mass, the axial-vector coupling constant,
and the pion-decay constant, respectively.

  In sec.\ \ref{elwpc} we saw that, in the purely mesonic sector,
contributions of $n$-loop diagrams are at least of order ${\cal O}(q^{2n+2})$,
i.e., they are suppressed by $q^{2n}$ in comparison with tree-level diagrams.
   An important ingredient in deriving this result was the fact that we
treated the squared pion mass as a small quantity of order $q^2$.
   Such an approach is motivated by the observation that the masses of the
Goldstone bosons must vanish in the chiral limit.
   In the framework of ordinary chiral perturbation theory $M_\pi^2\sim \hat m$
which translates into a momentum expansion of observables at fixed ratio $\hat
m/p^2$.
   On the other hand, there is no reason to believe that the masses of
hadrons other than the Goldstone bosons should vanish or become small in the
chiral limit.
   In other words, the nucleon mass entering the pion-nucleon Lagrangian
of eq.\ (\ref{lpin1}) should not be treated as a small quantity of, say, order
${\cal O}(q)$.
   Naturally the question arises how all this affects the calculation
of loop diagrams and the setup of a consistent power counting scheme.

   Our goal is to propose a renormalization procedure generating a power counting for
tree-level and loop diagrams of the (relativistic) EFT for baryons which is
analogous to that given in sec.\ \ref{elwpc} for mesons.
   Choosing a suitable renormalization condition will
allow us to apply the following power counting: a loop integration in $n$
dimensions counts as $q^n$, pion and fermion propagators count as $q^{-2}$ and
$q^{-1}$, respectively, vertices derived from ${\cal L}_{2k}$ and ${\cal L}_{\pi
N}^{(k)}$ count as $q^{2k}$ and $q^k$, respectively.
   Here, $q$ generically denotes a small expansion parameter such as,
e.g., the pion mass.
   In total this yields for the power $D$ of a diagram in the
one-nucleon sector the standard formula
\begin{eqnarray}
\label{dimension1} D&=&n N_L - 2 I_\pi - I_N +\sum_{k=1}^\infty 2k N^\pi_{2k}
+\sum_{k=1}^\infty k N_k^N\\
\label{dimension2} &=&1+(n-2)N_L+\sum_{k=1}^\infty 2(k-1) N^\pi_{2k}
+\sum_{k=1}^\infty (k-1) N_k^N\nonumber\\ &&\\&\geq&\mbox{1 in 4
dimensions},\nonumber
\end{eqnarray}
   where $N_L$, $I_\pi$, $I_N$, $N_{2k}^\pi$, and $N_k^N$ denote the
number of independent loop momenta, internal pion lines, internal nucleon lines,
vertices originating from ${\cal L}_{2k}$, and vertices originating from ${\cal
L}_{\pi N}^{(k)}$, respectively.

\begin{figure}
\begin{center}
\resizebox{0.2\textwidth}{!}{%
\includegraphics{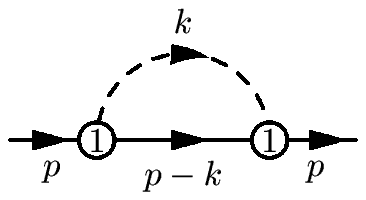}
}
\end{center}
\caption{One-loop contribution to the nucleon self-energy.
The number 1 in the interaction blobs refers to ${\cal L}_{\pi N}^{(1)}$.}
\label{fig:NukSE}
\end{figure}

  According to eq.\ (\ref{dimension2}), one-loop calculations in the
single-nucleon sector should start contributing at ${\cal O}(q^{n-1})$.
   For example, let us consider the one-loop contribution of fig.\ \ref{fig:NukSE}
to the nucleon self-energy.
   According to eq.\ (\ref{dimension1}), the renormalized result should
be of the order
\begin{equation}
\label{dexample} D=n\cdot 1-2\cdot 1-1\cdot 1+1\cdot 2=n-1.
\end{equation}
   We will see below that the corresponding renormalization scheme is
more complicated than in the mesonic sector.

   An explicit calculation yields \cite{Fuchs:2003qc}
\begin{eqnarray*}
\label{sigmaaresult} \lefteqn{\Sigma_{\rm loop}= -\frac{3
{\stackrel{\circ}{g_{A}}}_0^2}{4 F_0^2}\left\{
\vphantom{-\frac{(p^2-m^2)p\hspace{-.5em}/}{2p^2}}
(p\hspace{-.45em}/\hspace{.1em}+m)I_N
+M^2(p\hspace{-.45em}/\hspace{.1em}+m)I_{N\pi}(-p,0)\right.}\nonumber\\
&&\left. -\frac{(p^2-m^2)p\hspace{-.45em}/\hspace{.1em}}{2p^2}[
(p^2-m^2+M^2)I_{N\pi}(-p,0)+I_N-I_\pi]\right\},
\end{eqnarray*}
where the relevant loop integrals are defined as
\begin{eqnarray}
\label{Ipi}
I_\pi&=& \mu^{4-n}\int\frac{d^nk}{(2\pi)^n}\frac{i}{k^2-M^2+i0^+},\\
\label{IN}
I_N&=& \mu^{4-n}\int\frac{d^nk}{(2\pi)^n}\frac{i}{k^2-m^2+i0^+},\\
\label{INpi} I_{N\pi}(-p,0)&=&\mu^{4-n}\int\frac{d^nk}{(2\pi)^n}
\frac{i}{[(k-p)^2-m^2+i0^+]}\nonumber\\
&&\hspace{2em}\times\frac{1}{k^2-M^2+i0^+}.
\end{eqnarray}
   Applying the $\widetilde{\rm MS}$ renormalization scheme of ChPT
\cite{Gasser:1983yg,Gasser:1987rb}---indicated by ``r''---one obtains
\begin{displaymath}
\Sigma_{\rm loop}^r=-\frac{3 g_{Ar}^2}{4 F_r^2}\left[
-\frac{M^2}{16\pi^2}(p\hspace{-.4em}/\hspace{.1em}+m) +\cdots\right]= {\cal
O}(q^2),
\end{displaymath}
where $M^2$ is the lowest-order expression for the squared pion mass.
   In other words, the $\widetilde{\rm MS}$-renormalized result does not
produce the desired low-energy behavior of eq.\ (\ref{dexample}).
   This finding has widely been interpreted as the absence of a systematic
power counting in the relativistic formulation of ChPT.

\subsection{Heavy-baryon approach}
\label{hb}
   One possibility of overcoming the problem of power counting was provided
in terms of heavy-baryon chiral perturbation theory (HBChPT)
\cite{Jenkins:1990jv,Bernard:1992qa} resulting in a power counting scheme which
follows eqs.\ (\ref{dimension1}) and (\ref{dimension2}).
   The basic idea consists in dividing nucleon momenta into a large piece
close to on-shell kinematics and a soft residual contribution: $p = m v +k_p$,
$v^2=1$, $v^0\ge 1$ [often $v^\mu = (1,0,0,0)$].
   The relativistic nucleon field is expressed in terms of
velocity-dependent fields,
\begin{displaymath}
\Psi(x)=e^{-im v \cdot x} ({\cal N}_v +{\cal H}_v),
\end{displaymath}
with
\begin{displaymath}
{\cal N}_v=e^{+im v\cdot x}\frac{1}{2}(1+v\hspace{-.5em}/)\Psi,\quad {\cal
H}_v=e^{+im v\cdot x}\frac{1}{2}(1-v\hspace{-.5em}/)\Psi.
\end{displaymath}
   Using the equation of motion for ${\cal H}_v$, one can
eliminate ${\cal H}_v$ and obtain a Lagrangian for ${\cal N}_v$ which, to lowest
order, reads \cite{Bernard:1992qa}
\begin{displaymath}
\widehat{\cal L}^{(1)}_{\pi N}=\bar{\cal N}_v(iv\cdot D + g_A S_v\cdot u) {\cal
N}_v+{\cal O}(1/m).
\end{displaymath}
   The result of the heavy-baryon reduction is a $1/m$ expansion of the
Lagrangian similar to a Foldy-Wouthuysen expansion with a power counting along
eqs.\ (\ref{dimension1}) and (\ref{dimension2}).

\subsection{Pion electroproduction near threshold and the axial radius}
   As an example illustrating the strength of the EFT approach we consider pion electroproduction
   $\gamma^\ast(k)+N(p_i)\to \pi^i(q)+N(p_f)$ near threshold (for an overview,
see ref.~\cite{Drechsel:1992pn}) and the extraction of the nucleon axial radius.
   To that end we introduce the Green functions
\begin{eqnarray*}
\label{g1} {\cal M}^\mu_{A, i}&=& \langle N(p_f) | A^\mu_i(0) | N(p_i) \rangle,\\
\label{g2} {\cal M}_{JA, i}^{\mu\nu} &=&\int d\,^4 x\, e^{iq\cdot x} \langle
N(p_f) | T\left[J^\mu(0)
A^\nu_i(x)\right] | N(p_i) \rangle,\\
\label{g3} {\cal M}_{JP, i}^{\mu} &=& \int d\,^4 x\, \,e^{iq\cdot x} \langle
N(p_f) | T\left[J^\mu(0) P_i(x) \right] | N(p_i) \rangle,
\end{eqnarray*}
   where the subscripts $A$, $J$ and $P$ refer to {\em axial-vector current}, {\em
electromagnetic current} and {\em pseudoscalar density} and $i$ refers to the
$i$th isospin component of the axial-vector current or the pseudoscalar density,
respectively.
   The so-called  Adler-Gilman relation \cite{Adler:1966} provides the chiral Ward identity
\begin{equation}
\label{gfrel} q_\nu {\cal M}_{J A, i}^{\mu\nu}= i\hat{m}{\cal M}_{J P, i}^\mu
+\epsilon_{3ij}{\cal M}^\mu_{A,j}
\end{equation}
relating the three Green functions.
   In the one-photon-exchange approximation, the invariant amplitude
for pion electroproduction can be written as ${\cal M}_i=-ie \epsilon_\mu {\cal
M}^\mu_i$, where $\epsilon_\mu= e\bar{u}\gamma_\mu u/k^2$ is the polarization
vector of the virtual photon and ${\cal M}^\mu_i$ the transition-current matrix
element:
\begin{equation}
\label{mmu} {\cal M}^\mu_i=\langle N(p_f), \pi^i(q)|J^\mu(0)|N(p_i)\rangle.
\end{equation}
   The relation between the Adler-Gilman relation, eq.~(\ref{gfrel}), and pion electroproduction is
established in terms of the Lehmann-Symanzik-Zimmermann reduction formula,
 \begin{eqnarray*}
{\cal M}^\mu_i &=& -i\frac{\hat{m}}{M_\pi^2F_\pi}\lim_{q^2\to M_\pi^2} (q^2 -
M_\pi^2) {\cal M}_{J P, i}^\mu\\
         &=& \frac{1}{M_\pi^2 F_\pi}\lim_{q^2\to M_\pi^2}(q^2 - M_\pi^2)
(\epsilon_{3ij} {\cal M}^\mu_{A,j}-q_\nu {\cal M}^{\mu\nu}_{JA,i}).
\end{eqnarray*}
   At threshold, the center-of-mass transition current can be parameterized in
   terms of two s-wave amplitudes $E_{0+}$ and $L_{0+}$
\begin{displaymath}
\left.e\vec{M}\right|_{\rm thr}=\frac{4\pi W}{m_N}\left[i\vec\sigma_\perp
E_{0+}(k^2) +i\vec{\sigma}_\parallel L_{0+}(k^2)\right],
\end{displaymath}
where $W$ is the total center-of-mass energy, $\vec{\sigma}_\parallel=\vec
\sigma\cdot\hat k\hat k$ and $\vec\sigma_\perp= \vec
\sigma-\vec{\sigma}_\parallel$.

\begin{figure}
\begin{center}
\resizebox{0.45\textwidth}{!}{%
\includegraphics{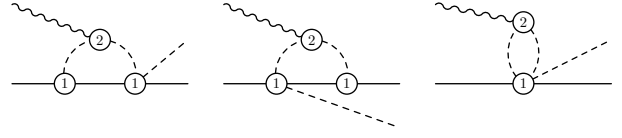}
}
\end{center}
\caption{One-loop contributions leading to a modification
of the $k^2$ dependence of $E_{0+}^{(-)}$.}
\label{fig:Bild4}
\end{figure}

   The contribution from pion loops (see fig.\ \ref{fig:Bild4}) has been analyzed in
\cite{Bernard:1992ys} and leads to a modification of the $k^2$ dependence of the
electric dipole amplitude $E_{0+}^{(-)}$ [at ${\cal O}(q^3)$]
\begin{eqnarray}
\label{e0plus} E^{(-)}_{0+}(k^2)&=&\frac{eg_A}{8\pi F_\pi}
\left[1+\frac{k^2}{4m^2_N}\left(\kappa_v+\frac{1}{2}\right)+\frac{k^2}{6}r^2_A\right.
\nonumber\\
&&+\left. \frac{M_\pi^2}{8 \pi^2 F^2_\pi}f\left(\frac{k^2}{M_\pi^2}\right)
+\cdots\right],
\end{eqnarray}
where $\kappa_v=3.706$ is the isovector anomalous magnetic moment of the nucleon
and $r_A$ is the axial radius.
   The first line corresponds to the traditional expression obtained in the
framework of the partially conserved axial-vector current hypothesis (see, e.g.,
\cite{Scherer:1991cy}).
   The second line generates the modification
$$\frac{M_\pi^2}{8 \pi^2 F^2_\pi}f\left(\frac{k^2}{M_\pi^2}\right)
=\frac{k^2}{128 F^2_\pi}\left(1-\frac{12}{\pi^2}\right)+\cdots.
$$

   The reaction $p(e,e'\pi^+)n$ has been measured at MAMI at an invariant mass
of $W=1125$ MeV (corresponding to a pion center of mass momentum of
$|\vec{q}^\ast|=112$ MeV) and photon four-momentum transfers of $Q^2=0.117,
0.195$ and 0.273 GeV$^2$ \cite{Liesenfeld:1999mv}.
   Using an effective-Lagrangian model and a dipole form as an ansatz for the axial
form factor $G_A$, an axial mass of
\begin{displaymath}
\tilde{M}_A=(1.077\pm 0.039)\,\mbox{GeV}
\end{displaymath}
was extracted which has to be compared with the average of neutrino scattering
experiments
\begin{displaymath}
M_A=(1.026\pm 0.021)\,\mbox{GeV}.
\end{displaymath}
   Defining $\tilde{M}_A=M_A+\Delta M_A$, the difference between the two results can nicely
be explained in terms of the additional $k^2$ dependence of eq.~(\ref{e0plus})
yielding $\Delta M_A=0.056$ GeV.
   In the meantime, the experiment has been repeated including an additional
value of $Q^2=0.058$ GeV$^2$ \cite{Baumann:2004} and is currently being analyzed.

   Recently, there have been claims that pion electroproduction data at threshold cannot be
interpreted in terms of $G_A$ \cite{Haberzettl:2000sm}.
   However, as was shown in \cite{Fuchs:2003vw}, using minimal coupling alone does
not respect the constraints due to chiral symmetry.
   In the framework of the most general Lagrangian, this can be seen by
considering the $b_{23}$ term of the ${\cal O}(q^3)$ Lagrangian
\cite{Ecker:1995rk},
\begin{equation}
\label{lb23} {\cal L}_{\rm eff}^{(3)} = \frac{1}{2(4\pi F)^2}b_{23}
\,\bar{\Psi}\gamma^\mu\gamma_5 [D^\nu, f_{-\mu\nu}]\Psi+\cdots
\end{equation}
with
\begin{eqnarray*}
  f_{-\mu\nu}
   &=& -2(\partial_\mu a_\nu - \partial_\nu a_\mu) + 2i \left(
       [v_\mu, a_\nu] - [v_\nu, a_\mu] \right) \\
   && +\frac{i}{F} \left[\vec{\tau}\cdot\vec{\pi},
      \partial_\mu v_\nu - \partial_\nu v_\mu
   \right]
+\cdots.
\end{eqnarray*}
   The Lagrangian of eq.~(\ref{lb23}) is of a non-minimal type and the
three terms contribute to the axial-vector matrix element, the $J A$ Green
function and pion electroproduction relevant to the Adler-Gilman relation.
   As a result it was confirmed that threshold pion electroproduction is
indeed a tool to obtain information on the axial form factor of the nucleon (see
\cite{Fuchs:2003vw} for details).

\subsection{Virtual Compton scattering and generalized polarizabilities}

   As a second example, let us discuss the application of HBChPT to the
calculation of the so-called generalized polarizabilities
\cite{Arenhovel:1974,Guichon:1995pu}.
   The virtual Compton scattering (VCS) amplitude
$T_{\rm VCS}$ is accessible in the reaction $e^-p\to e^-p\gamma$.
   Model-independent predictions,
based on Lorentz invariance, gauge invariance, crossing symmetry, and the
discrete symmetries, have been derived in ref.\ \cite{Scherer:1996ux}.
  Up to and including terms of second order in the momenta $|\vec{q}\,|$ and
$|\vec{q}\,'|$ of the virtual initial and real final photons, the amplitude is
completely specified in terms of quantities which can be obtained from elastic
electron-proton scattering and real Compton scattering, namely $m_N$, $\kappa$,
$G_E$, $G_M$, $r^2_E$, $\alpha_p$ and $\beta_p$.
   The generalized polarizabilities
(GPs) of ref.\ \cite{Guichon:1995pu} result from an analysis of the residual
piece in terms of electromagnetic multipoles.
   A restriction to the lowest-order, i.e.\ linear terms in $\omega'$ leads to
only electric and magnetic dipole radiation in the final state.
   Parity and angular-momentum selection rules, charge-conjugation symmetry,
and particle crossing generate six independent GPs
\cite{Guichon:1995pu,Drechsel:1996ag,Drechsel:1997xv}.

\begin{figure}
\begin{center}
\resizebox{0.33\textwidth}{!}{%
\includegraphics{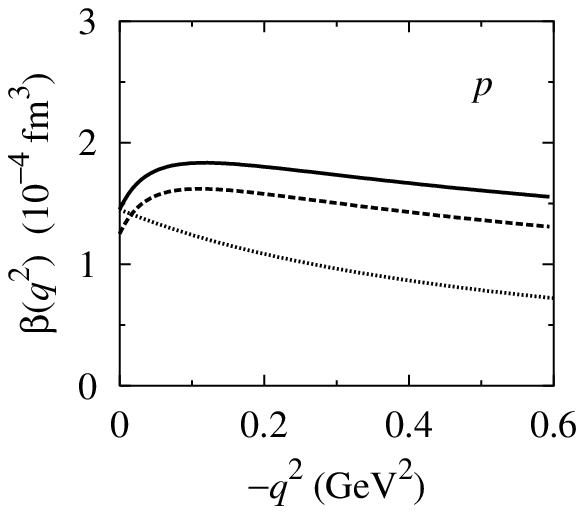}
}
\resizebox{0.33\textwidth}{!}{%
\includegraphics{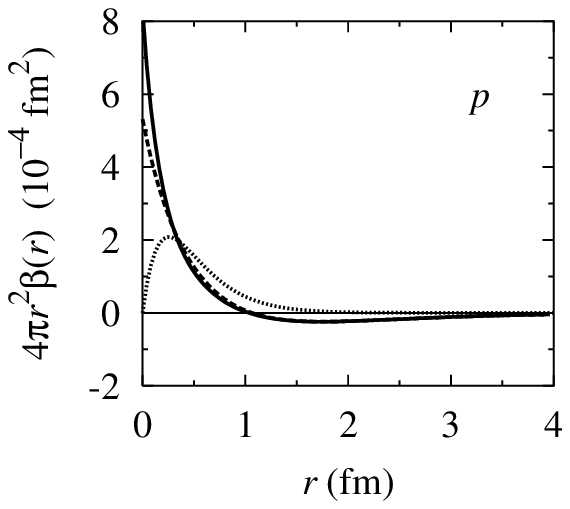}
}
\end{center}
\caption{Generalized
magnetic polarizability $\beta(q^2)$ and density of magnetic polarizability
$\beta(r)$ for the proton. Dashed lines: contribution of pion loops; solid lines:
total contribution; dotted lines: VMD predictions normalized to ${\beta}(0)$
\cite{L'vov:2001fz}.} \label{fig:beta}
\end{figure}

   The first results for the two structure functions $P_{LL}-P_{TT}/\epsilon$
and $P_{LT}$ at $Q^2=0.33$ GeV$^2$ were obtained from a dedicated VCS experiment
at MAMI \cite{Roche:2000ng}.
   Results at higher four-momentum transfer squared
$Q^2=0.92$ and $Q^2=1.76$ GeV$^2$ have been reported in ref.\
\cite{Laveissiere:2004nf}.
   Additional data are expected from MIT/Bates for $Q^2 =
0.05$ GeV$^2$ aiming at an extraction of the magnetic polarizability.
   Moreover, data in the resonance region have been taken at JLab
for $Q^2=1$ GeV$^2$ \cite{Fonvieille:2004rb} which have been analyzed in the
framework of the dispersion relation formalism of ref.\
\cite{Pasquini:2001yy,Drechsel:2002ar}.
   Table \ref{H1:b2:tableresults} shows the experimental results of
\cite{Roche:2000ng} in combination with various model calculations.
   Clearly, the experimental precision of \cite{Roche:2000ng} already
allows for a critical test of the different models.
   Within ChPT and the linear sigma model, the GPs are essentially
due to pionic degrees of freedom.
   Due to the small pion mass the effect in the spatial distributions
extends to larger distances (see also fig.\ \ref{fig:beta}).
   On the other hand, the constituent quark model and other phenomenological
models involving Gau{\ss} or dipole form factors typically show a faster decrease in
the range $Q^2 < 1$ GeV$^2$.

\begin{table*}
\begin{center}
\caption{
Experimental results and theoretical predictions for the structure functions
$P_{LL}-P_{TT}/\epsilon$ and $P_{LT}$ at $Q^2=0.33$ GeV$^2$ and $\epsilon=0.62$.
 $\ast$ makes use of symmetry under particle crossing and charge conjugation
which is not a symmetry of the nonrelativistic quark model.}
\label{H1:b2:tableresults}
\begin{tabular}{ccc}
\hline\noalign{\smallskip}  &$P_{LL}-P_{TT}/\epsilon$ $[\mbox{GeV}^{-2}]$ &
$P_{LT}$ $[\mbox{GeV}^{-2}]$\\
\noalign{\smallskip}\hline\noalign{\smallskip} Experiment \cite{Roche:2000ng} &
$23.7\pm 2.2_{\rm stat.}\pm 4.3_{\rm syst.} \pm 0.6_{\rm syst.norm.}$ &
 $-5.0
\pm 0.8_{\rm stat.} \pm 1.4_{\rm syst.} \pm 1.1_{\rm syst. norm.}$
\\
Linear sigma model \cite{Metz:1996fn} & 11.5 & 0.0\\
Effective Lagrangian model \cite{Vanderhaeghen:1996iz} & 5.9 & $-1.9$\\

HBChPT \cite{Hemmert:1997at} & 26.0 & $-5.3$\\

Nonrelativistic quark model \cite{Pasquini:2000ue} & $19.2|14.9^\ast$ & $-3.2|-4.5^\ast$\\
\noalign{\smallskip}\hline
\end{tabular}
\end{center}
\end{table*}

\begin{figure}
\begin{center}
\resizebox{0.35\textwidth}{!}{%
\includegraphics{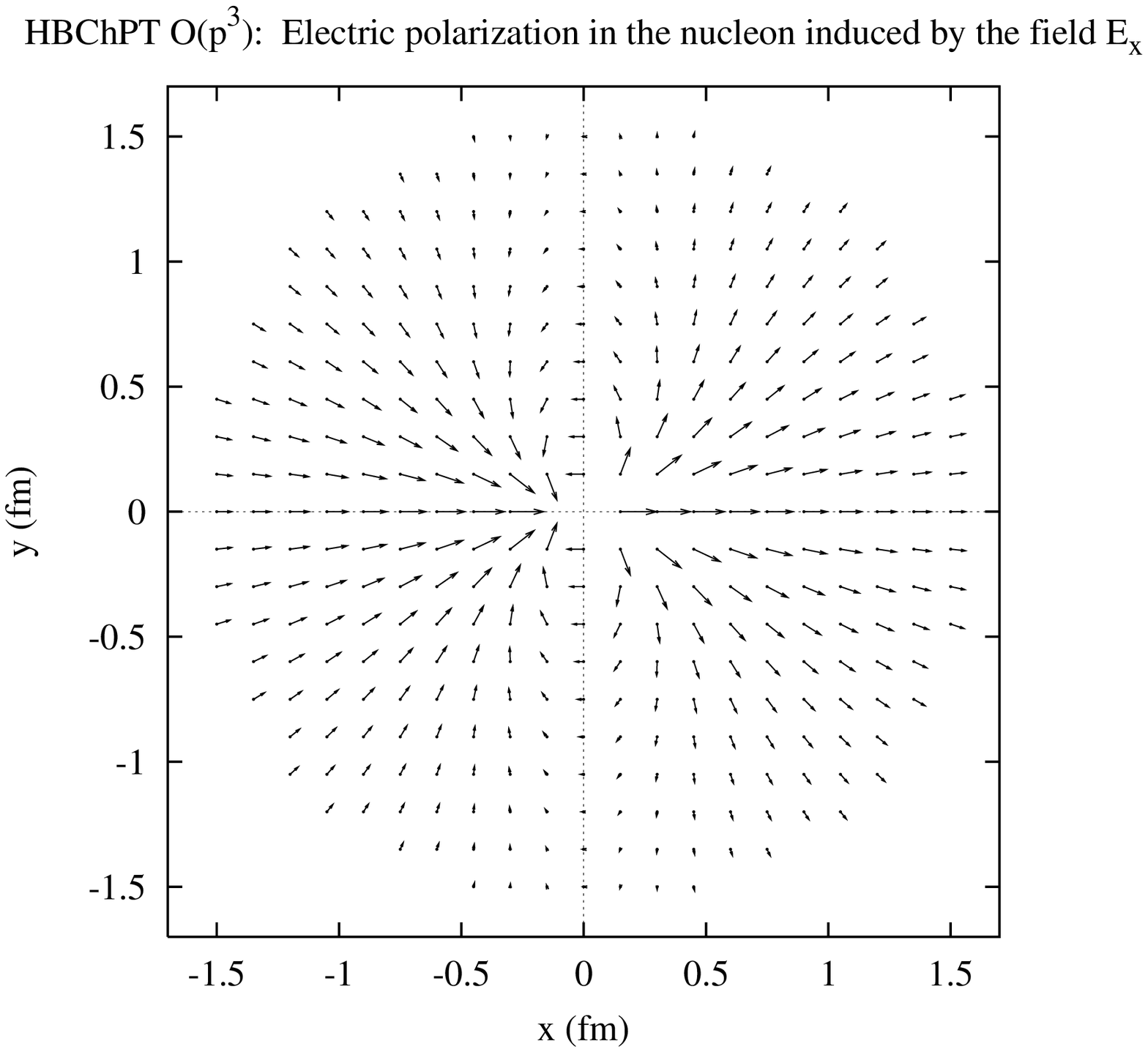}
}
\end{center}
\caption{Scaled electric polarization
$r^3 \alpha_{i1}$ [10$^{-3}$ fm$^3$] \cite{Lvov:2004}. The applied electric field
points in the $x$ direction.} \label{fig:polarization}
\end{figure}

   A covariant definition of the spin-averaged dipole polarizabilities has been
proposed in ref.~\cite{L'vov:2001fz}.
   It was shown that {\em three} generalized dipole polarizabilities are needed to reconstruct
spatial distributions.
   For example, if the nucleon is exposed to a static and uniform external
electric field  $\vec{E}$, an electric polarization $\vec{\cal P}$ is generated
which is related to the {\em density} of the induced electric dipole moments,
\begin{equation}
\label{H1:b:d-induced} {\cal P}_i(\vec r) = 4\pi\alpha_{ij}(\vec r)\,E_j.
\end{equation}
   The tensor $\alpha_{ij}(\vec r)$,  i.e.~the density of the full electric
polarizability of the system, can be expressed as \cite{L'vov:2001fz}
\begin{eqnarray*}
\alpha_{ij}(\vec r) &=&
     \alpha_L(r) \hat r_i \hat r_j
      + \alpha_T(r) (\delta_{ij} - \hat r_i \hat r_j)\\
     && + \frac{3\hat r_i \hat r_j - \delta_{ij}}{r^3}
        \int_r^\infty [\alpha_L(r')-\alpha_T(r')]\,r'^2\,dr',
\end{eqnarray*}
   where $\alpha_L(r)$ and $\alpha_T(r)$ are Fourier transforms
of the generalized longitudinal and transverse electric polarizabilities
$\alpha_L(q)$ and $\alpha_T(q)$, respectively.
    In particular, it is important to realize that both longitudinal and
transverse polarizabilities are needed to fully recover the electric polarization
$\vec{\cal P}$.
   Figure \ref{fig:polarization} shows the induced polarization inside a proton
as calculated in the framework of HBChPT at ${\cal O}(q^3)$ \cite{Lvov:2004} and
clearly shows that the polarization, in general, does not point into the
direction of the applied electric field.

   Similar considerations apply to an external magnetic field.
   Since the magnetic induction is always transverse
(i.e., $\vec\nabla\cdot\vec B=0$), it is sufficient to consider $\beta_{ij}(\vec
r)=\beta(r)\delta_{ij}$ \cite{L'vov:2001fz}.
   The induced magnetization $\vec{\cal
M}$ is given in terms of the density of the magnetic polarizability as $\vec{\cal
M}(\vec r) = 4\pi\beta(r)\vec B$ (see fig.\ \ref{fig:beta}).

\subsection{Manifestly Lorentz-invariant baryon chiral perturbation theory}
   Unfortunately, when considering higher orders in the chiral expansion, the expressions due
to $1/m$ corrections of the Lagrangian become increasingly complicated.
   Secondly, not all of the scattering amplitudes, evaluated perturbatively
in the heavy-baryon framework, show the correct analytical behavior in the
low-energy region.
   Finally, with an increasing complexity of processes, the use of computer
algebra systems becomes almost mandatory.
   The relevant techniques have been developed for calculations in the Standard Model and thus
refer to loop integrals of the manifestly Lorentz-invariant type.

   In the following we will concentrate on one of
several methods that have been suggested to obtain a consistent power counting in
a manifestly Lorentz-invariant approach
\cite{Tang:1996ca,Ellis:1997kc,Becher:1999he,Lutz:1999yr,%
Gegelia:1999gf,Gegelia:1999qt,Lutz:2001yb,Fuchs:2003qc},
   namely, the so-called extended on-mass-shell (EOMS)
renormalization scheme \cite{Fuchs:2003qc}.
   The central idea of the EOMS scheme consists of performing additional
subtractions beyond the $\widetilde{\rm MS}$ scheme.
   Since the terms violating the power counting are analytic in small
quantities, they can be absorbed by counterterm contributions.
   Let us illustrate the approach in terms of the integral
\begin{displaymath}
H(p^2,m^2;n)= \int \frac{d^n k}{(2\pi)^n} \frac{i}{[(k-p)^2-m^2+i0^+][k^2+i0^+]},
\end{displaymath}
where $\Delta=(p^2-m^2)/m^2={\cal O}(q)$ is a small quantity.
   We want the (renormalized) integral to be of the order $D=n-1-2=n-3$.
   Applying the dimensional counting analysis of ref.\ \cite{Gegelia:1994zz}
(for an illustration, see the appendix of ref.\ \cite{Schindler:2003je}), the
result of the integration is of the form \cite{Fuchs:2003qc}
\begin{displaymath}
H\sim F(n,\Delta)+\Delta^{n-3}G(n,\Delta),
\end{displaymath}
where $F$ and $G$ are hypergeometric functions and are analytic in $\Delta$ for
any $n$.
   Hence, the part containing $G$ for noninteger $n$ is proportional to
a noninteger power of $\Delta$ and satisfies the power counting.
   On the other hand $F$ violates the power counting.
   The crucial observation is that the part proportional to $F$ can be
obtained by {\em first} expanding the integrand in small quantities and {\em
then} performing the integration for each term \cite{Gegelia:1994zz}.
   This observation suggests the following procedure: expand the integrand in
small quantities and subtract those (integrated) terms whose order is smaller
than suggested by the power counting.
   In the present case, the subtraction term reads
\begin{displaymath}
H^{\rm subtr}=\int \frac{d^n k}{(2\pi)^n}\left. \frac{i}{[k^2-2p\cdot k
+i0^+][k^2+i0^+]}\right|_{p^2=m^2}
\end{displaymath}
and the renormalized integral is written as $ H^R=H-H^{\rm subtr}={\cal O}(q) $
as $n\to 4$.
   In the infrared renormalization (IR) scheme of Becher and Leutwyler
\cite{Becher:1999he}, one would keep the contribution proportional to $G$ (with
subtracted divergences when $n$ approaches 4) and completely drop the $F$ term.

   Let us conclude this section with a few remarks.
   With a suitable renormalization condition one can also obtain a consistent power counting in
manifestly Lorentz-invariant baryon chiral perturbation theory including, e.g.,
vector mesons \cite{Fuchs:2003sh} or the $\Delta(1232)$ resonance
\cite{Hacker:2005fh} as explicit degrees of freedom. Secondly, the infrared
regularization of Becher and Leutwyler \cite{Becher:1999he} may be formulated in
a form analogous to the EOMS renormalization \cite{Schindler:2003xv}.
   Finally, using a toy model we have explicitly demonstrated the application of both
infrared and extended on-mass-shell renormalization schemes to multiloop diagrams
by considering as an example a two-loop self-energy diagram
\cite{Schindler:2003je}.
   In both cases the renormalized diagrams satisfy a
straightforward power counting.

\subsection{Applications}

\begin{figure}
\begin{center}
\resizebox{0.45\textwidth}{!}{%
\includegraphics{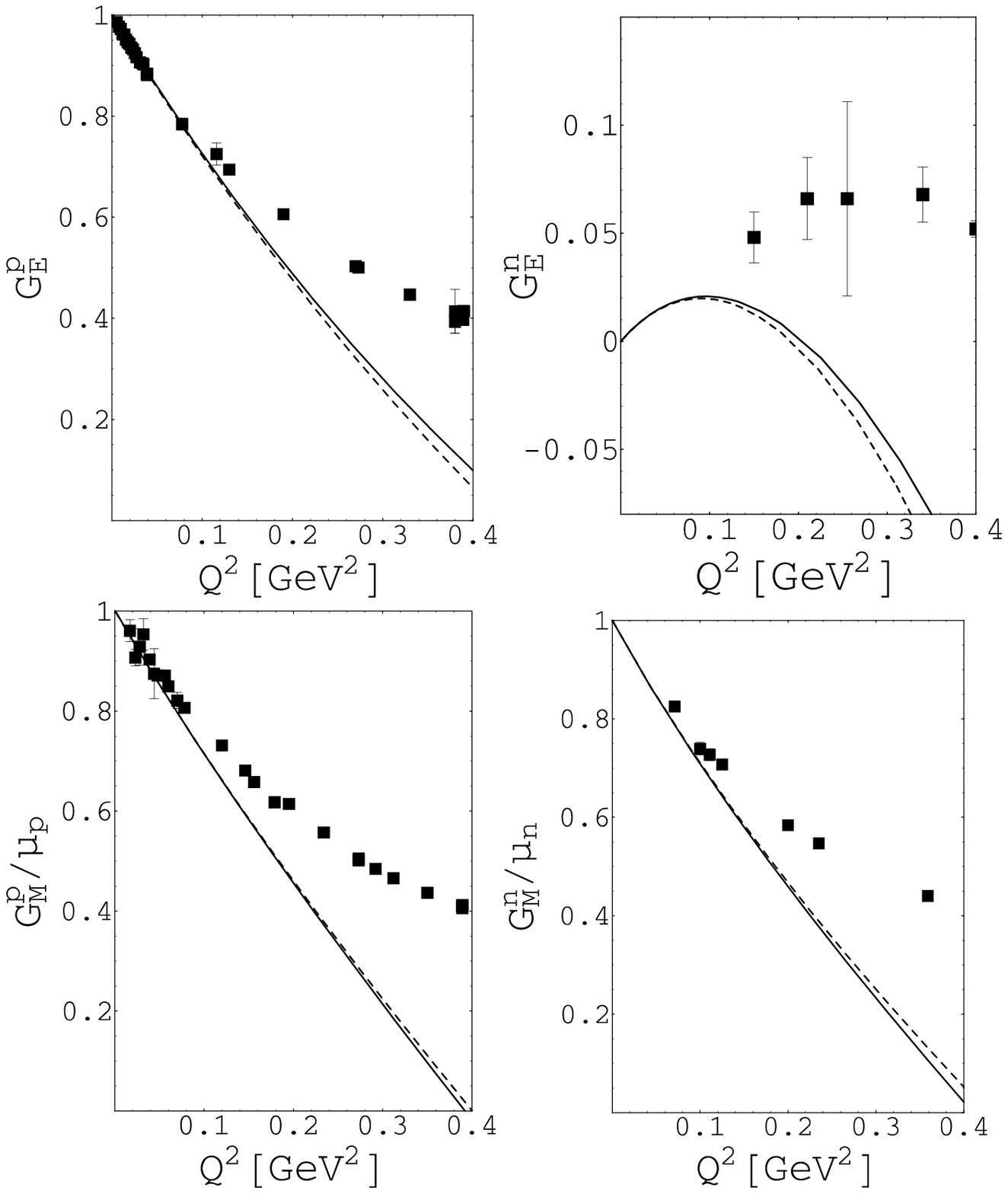}
}
\end{center}
\caption{ The Sachs form factors of the nucleon in
manifestly Lorentz-invariant chiral perturbation theory at ${\cal O}(q^4)$
without vector mesons. Full lines: results in the extended on-mass-shell scheme;
dashed lines: results in infrared regularization. The experimental data are taken
from ref.\ \cite{Friedrich:2003iz}.} \label{fig:G_ohne}
\end{figure}

   The EOMS scheme has been applied in several calculations such
as the chiral expansion of the nucleon mass, the pion-nucleon sigma term, and the
scalar form factor \cite{Fuchs:2003kq}, the masses of the ground-state baryon
octet \cite{Lehnhart:2004vi} and the nucleon electromagnetic form factors
\cite{Fuchs:2003ir,Schindler:2005ke}.

 As an example, let us here consider the electromagnetic form factors
of the nucleon which are defined via the matrix element of the electromagnetic
current operator as
\begin{eqnarray*}
\lefteqn{\langle N(p_f)\left| J^\mu(0) \right| N(p_i) \rangle=}\\
&&   \bar{u}(p_f)\left[\gamma^\mu F_1^N(Q^2)+
\frac{i\sigma^{\mu\nu}q_\nu}{2m_N}F_2^N(Q^2) \right] u(p_i), \,\, N=p,n,
\end{eqnarray*}
where $q=p_f-p_i$ is the momentum transfer and $Q^2\equiv-q^2=-t \ge 0$.
   Figure \ref{fig:G_ohne} shows the results for the electric and
magnetic Sachs form factors $G_E=F_1 - Q^2/(4m_N^2) F_2$ and $G_M= F_1 + F_2$ at
${\cal O}(q^4)$ in the momentum transfer region $0\,{\rm GeV^2}\le Q^2\le
0.4\,{\rm GeV^2}$ without explicit vector-meson degrees of freedom
\cite{Fuchs:2003ir}.
    The ${\cal O}(q^4)$ results only provide a decent description
up to $Q^2=0.1\,\mbox{GeV}^2$ and do not generate sufficient curvature for larger
values of $Q^2$.
   The perturbation series converges, at best, slowly and
higher-order contributions must play an important role.

\begin{figure}
\begin{center}
\resizebox{0.4\textwidth}{!}{%
\includegraphics{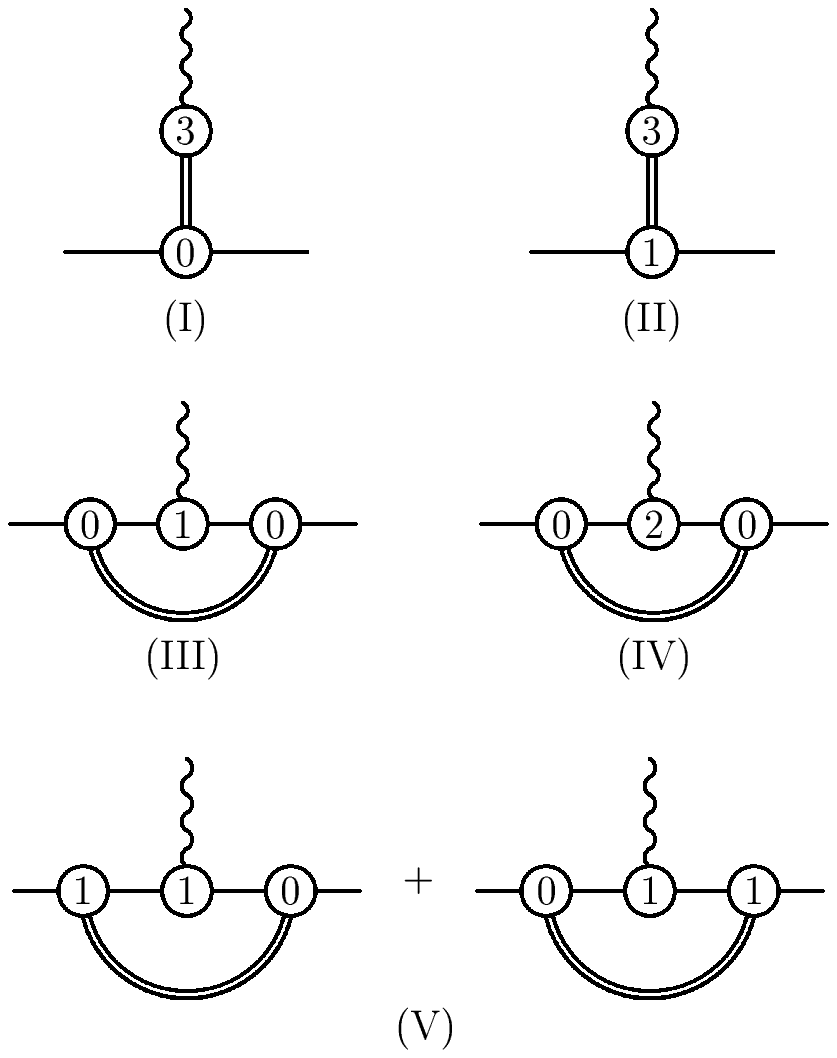}
}
\end{center}
\caption{Feynman diagrams involving vector mesons (double lines)
contributing to the electromagnetic form factors up to and including ${\cal
O}(q^4)$.} \label{fig:Dia_mit}
\end{figure}

\begin{figure}
\begin{center}
\resizebox{0.45\textwidth}{!}{%
\includegraphics{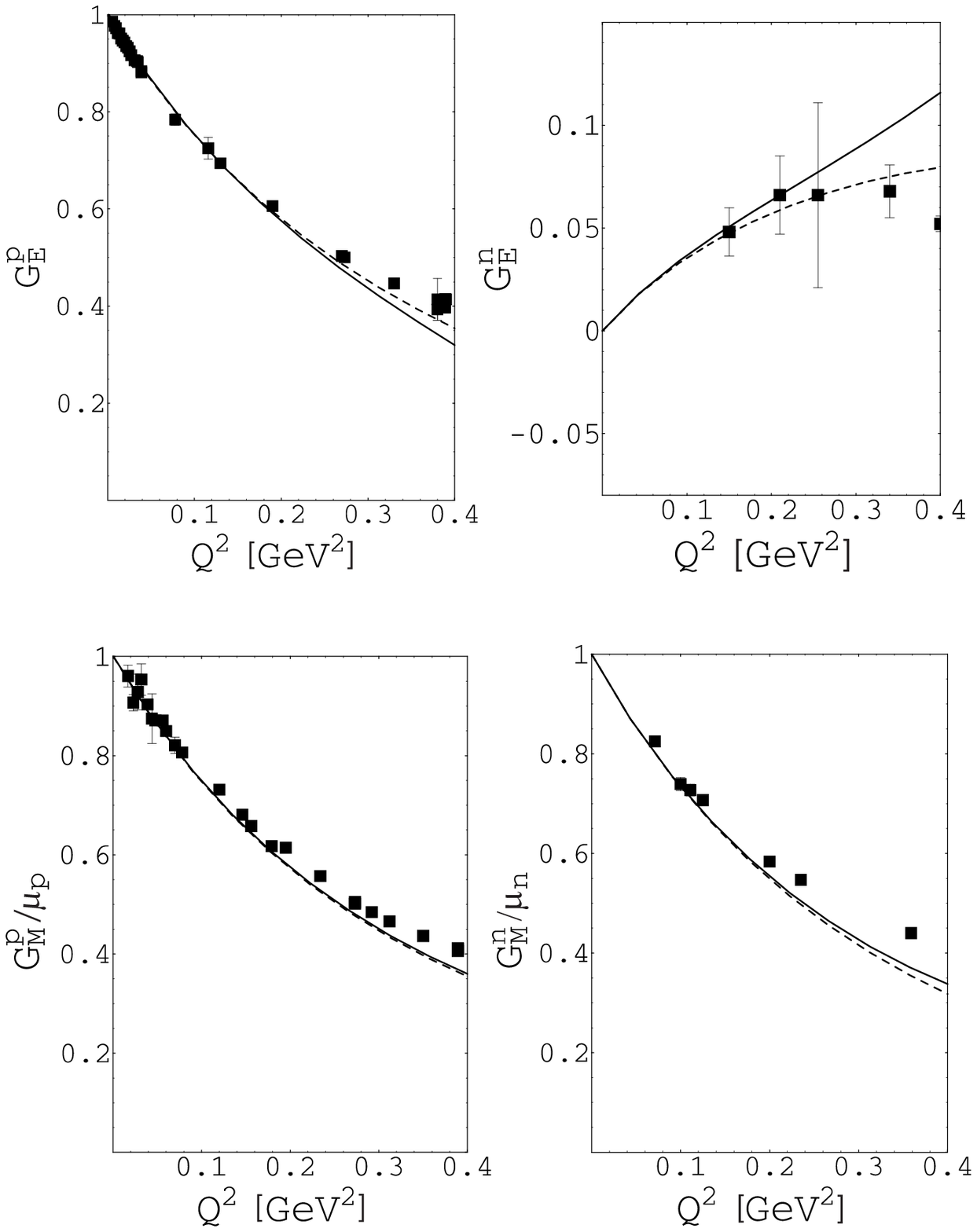}
}
\end{center}
\caption{\label{G_neu} The Sachs form factors of the nucleon in
manifestly Lorentz-invariant chiral perturbation theory at ${\cal O}(q^4)$
including vector mesons as explicit degrees of freedom. Full lines: results in
the extended on-mass-shell scheme; dashed lines: results in infrared
regularization. The experimental data are taken from ref.\
\cite{Friedrich:2003iz}.} \label{fig:G_neu}
\end{figure}

   Including the vector-meson degrees of
freedom along the lines of refs.~\cite{Fuchs:2003sh,Schindler:2003xv} generates
the additional diagrams of fig.~\ref{fig:Dia_mit}.
   The results for the Sachs form factors including vector-meson degrees of
freedom are shown in fig.~\ref{fig:G_neu}.
   As expected on phenomenological grounds \cite{Kubis:2000zd}, the quantitative
description of the data has improved considerably for $Q^2\ge 0.1$ GeV$^2$.
   The small difference between the two renormalization schemes is due to the
way how the regular higher-order terms of loop integrals are treated.
   Note that on an absolute scale the differences between the two schemes are
comparable for both $G_E^p$ and $G_E^n$.
   Numerically, the results are similar to those of ref.\ \cite{Kubis:2000zd}.
   Due to the renormalization condition, the contribution of the vector-meson
loop diagrams either vanishes (infrared renormalization scheme) or turns out to
be small (EOMS).
   Thus, in hindsight our approach puts the traditional phenomenological
vector-meson dominance model on a more solid theoretical basis.

\section{Summary}
   Chiral perturbation theory is a cornerstone of our understanding of
the strong interactions at low energies.
   Mesonic chiral perturbation theory has been tremendously successful and may be considered
as a full-grown and mature area of low-energy particle physics.
   The apparent conflict between the determination of the ${\cal O}(q^4)$
low-energy constants $(\bar l_6-\bar l_5)$ from radiative pion beta decay on the
one hand and the polarizability measurement on the other hand certainly requires
additional work, in particular, from the theoretical side.

   The impact on baryonic chiral perturbation theory due to the investigation of electromagnetic
reactions at MAMI such as elastic electron-nucleon scattering, (virtual) Compton
scattering and the electromagnetic production of pions cannot be overestimated.
   The possibility of a consistent manifestly Lorentz-invariant approach in
   combination with the rigorous inclusion of (axial-) vector-meson degrees of freedom
and of the $\Delta(1232)$ resonance open the door to an application of ChPT in an
extended kinematic region.

I would like to thank the organizers---Hartmuth\linebreak
 Arenh\"ovel, Hartmut
Backe, Dieter Drechsel, J\"org Friedrich, Karl-Heinz Kaiser and Thomas
Walcher---of the symposium {\em 20 Years of Physics at the Mainz Microtron MAMI}
and express my best wishes for the future.

\end{document}